\documentclass[runningheads,a4paper]{llncs}
\usepackage{amssymb,amsmath}
\usepackage{graphicx,tabularx,amsmath,amssymb,float,siunitx,soul,verbatim,dsfont}
\usepackage{setspace} 
\usepackage{comment}
\usepackage[nolist]{acronym}
\usepackage{multicol}
\usepackage{booktabs}
\usepackage{multirow}
\usepackage{caption}
\usepackage{subcaption}
\usepackage[hyphens]{url}
\usepackage{rotating}
\usepackage{pgf}
\usepackage{pgfplots}
\pgfplotsset{compat=1.11}
\usepackage{url}
\usepackage{colortbl}
\definecolor{Gray}{gray}{0.95}

\usepackage{soul}
\usepackage[normalem]{ulem}

\usepackage[textsize=footnotesize]{todonotes}

\newcommand{\iMP}[1]{\todo[color=yellow!10, inline]{\textbf{MP}: #1}}
\newcommand{\mc}{\mathcal}

\newcommand{\users}{\mc{U}}

\newcommand{\levels}{\mc{L}}
\newcommand{\safeguards}{\mc{S}}

\newcommand{\mixed}{(\mixdef,\mixatt)}
\newcommand{\BR}{\mathrm{BR}}
\newcommand{\mixedNE}{\mixdef^{\NE},\mixatt^{\NE}}
\newcommand{\NE}{\mathrm{NE}}
\newcommand{\zs}{\Gamma_0}

\newcommand{\plan}{p_{cj}}

\newcommand{\game}{\Gamma_{\sigma,\lambda}}

\newcommand{\equilibrium}{\mixdef^{NE}_{\sigma,\lambda}}

\newcommand{\mixdef}{\vec{\delta}}
\newcommand{\mixatt}{\vec{\alpha}}

\newcommand{\pair}{\sigma,\lambda}

\newif\ifExtended
\Extendedfalse

\begin{document}

\mainmatter  % start of an individual contribution

\title{Optimizing Investments in Cyber Hygiene \\ for Protecting Healthcare Users}
% \title{Optimizing Cyber Hygiene for Healthcare Safeguards: A Case Study for
% Security Awareness and Training in Healthcare}
\titlerunning{}
\author{
    Sakshyam Panda\inst{1} \and
    Emmanouil Panaousis\inst{2} \and \\
    George Loukas\inst{2} \and
    Christos Laoudias\inst{3} 
}
\authorrunning{Panda et al.}
% First names are abbreviated in the running head.
% If there are more than two authors, 'et al.' is used.
%
\institute
 {
    University of Surrey, UK\\
        \email{s.panda@surrey.ac.uk }
    \and University of Greenwich, UK \\
        \email{\{e.panaousis,g.loukas\}@greenwich.ac.uk }\\
    \and University of Cyprus, Cyprus \\
        \email{laoudias.christos@ucy.ac.cy}
    % \email{
    %     s.panda@surrey.ac.uk \\
    %     \{e.panaousis,g.loukas\}@greenwich.ac.uk \\
    %     laoudias.christos@ucy.ac.cy \\
    %     }
}

\maketitle              % typeset the header of the contribution

\begin{abstract}

Cyber hygiene measures are often recommended for strengthening an organization's security posture, especially for protecting against social engineering attacks that target the human element. However, the related recommendations are typically the same for all organizations and their employees, regardless of the nature and the level of risk for different groups of users. Building upon an existing cybersecurity investment model, this paper presents a tool for optimal selection of cyber hygiene safeguards, which we refer as the Optimal Safeguards Tool (OST). The model combines game theory and combinatorial optimization (0-1 Knapsack) taking into account the probability of each user group to being attacked, the value of assets accessible by each group, and the efficacy of each control for a particular group. The model considers indirect cost as the time employees could require for learning and trainning against an implemented control. Utilizing a game-theoretic framework to support the Knapsack optimization problem permits us to optimally select safeguards' application levels minimizing the aggregated expected damage within a security investment budget. 

We evaluate OST in a healthcare domain use case. In particular, on the Critical Internet Security (CIS) Control group 17 for implementing security awareness and training programs for employees belonging to the ICT, clinical and administration personnel of a hospital. We compare the strategies implemented by OST against alternative common-sense defending approaches for three different types of attackers: Nash, Weighted and Opportunistic. Our results show that Nash defending strategies are consistently better than the competing strategies for all attacker types with a minor exception where the Nash defending strategy, for a specific game, performs at least as good as other common-sense approaches. Finally, we illustrate the alternative investment strategies on different Nash equilibria (called plans) and discuss the optimal choice using the framework of  0-1 Knapsack optimization.

% The Defender is a decision maker, such as a security manager seeking 
% optimal decisions on the intensity, called \textit{level}, of each safeguard implementing
% Control 17.
% On the other hand, the Attacker is an adversary who is targeting different 
% individuals to gain access to an unauthorized area of the Defender's digital environments, 
% for example through spear phishing. In our evaluations, we consider three different 
% types of users that exhibit similar probabilities in getting attacked and compromised as well
% as total asset value they have access to. \MP{and how many CIS sub-controls?}
% Results show that \hl{...} \MP{we need some nice conclusions here based on comparisons with other
% non-game theoretic approaches.}

\end{abstract}

% 
% Introduction
%
\section{Introduction}

%Cyber security in healthcare - in general 

In the last few years, several cybersecurity incidents have taken place   
in the healthcare sector, including the WannaCry ransomware, which influenced globally the cybersecurity landscape\footnote{\scriptsize{\url{https://www.telegraph.co.uk/technology/2018/10/11/wannacry-cyber-attack-cost-nhs-92m-19000-appointments-cancelled.}}}. The 2018 Ponemon Cost of a Data Breach study\footnote{\scriptsize{\url{https://securityintelligence.com/series/ponemon-institute-cost-of-a-data-breach-2018.}}} shows that the healthcare industry has the
highest cost per record breached in a cyber incident, at \$408. This is almost twice the equivalent cost per record breached in the financial sector. This calls for the effective preparation of healthcare organizations in an ever-evolving cyber attack landscape. An example project that is addressing this from the perspective of training the users in the sector is H2020 CUREX 
project\footnote{\scriptsize{\url{https://cordis.europa.eu/project/rcn/220350/factsheet/en.}}}, which allows a healthcare provider to assess the realistic cybersecurity and privacy risks 
they are exposed to \cite{mohammadi2019curex}. 

Yet, a recent report from Mckinsey\footnote{\scriptsize{\url{https://www.mckinsey.com/business-functions/risk/our-insights/cyber-risk-measurement-and-the-holistic-cybersecurity-approach.}}} states that
almost all companies systematically over-invest in the protection of assets
that have no risk while at the same time they under-fund the protection of high-risk assets. 
Furthermore, regarding bearing costs of cybersecurity controls, in a survey from 
KPMG\footnote{\scriptsize{\url{https://advisory.kpmg.us/content/dam/advisory/en/pdfs/cyber-report-healthcare.pdf.}}}, 43\% of correspondents stated that they did not increase their cybersecurity budget even though high profile security breaches have been widely known. So, effective risk management is not only about assessing the risk correctly but also about selecting the controls that are optimal given the cost constraints of adopting them. To address the challenge of optimal control selection, in this paper, we formulate a model and tool for suggesting mathematically optimal \textit{cyber hygiene} strategies minimising the cyber risk.

Regarding cyber hygiene, we adopt the recent definition proposed by \cite{vishwanath2019cyber}, which relates it to ``\textit{the cyber security practices that online consumers should 
engage in to protect the safety and integrity of their personal information on their 
Internet enabled devices from being compromised in a cyber-attack.}''
Towards the goal of optimizing cyber hygiene, we extend the model presented in \cite{fielder2016decision} so that:
\begin{itemize}
    \item the Attacker's target is a user group (focusing on social engineering attacks) 
    instead of (asset, vulnerability) pair of the system;
    \item the Indirect cost of a safeguards' application depends not only on the 
    safeguard itself but also on the size of the user group (i.e., number of users) 
    and more specifically it increases with the group size;
    \item we adopt an aggregated risk model, as the objective function of Knapsack optimization
    problem, rather than the weakest-link defending 
    against a variety of attacks that can cause, in total, highest aggregated damage and;
    \item we use a ``small'' healthcare case study as a preliminary example to evaluate the OST against other common-sense approaches for a number of attacking strategies that have not been
    simulated in \cite{fielder2016decision}.
\end{itemize}

Our analysis results show that the game-theoretic approach increases risk control
efficacy, by selecting an optimal combination of safeguard application levels, compared
with alternative common-sense approaches. In addition, our use case designed for the healthcare domain exhibits a number of interchangeably optimal investment strategies subject to a budget constraint under the framework of 0-1 Knapsack optimization. 

The remainder of this paper is organized as follows. Section \ref{sec:relatedwork} presents
the related work in both the fields of (i) user-oriented cybersecurity safeguards and (ii) 
optimization of cybersecurity countermeasures including security investments. 
Section \ref{sec:model} presents both the game-theoretic model used to determine optimal cybersecurity safeguard plans as well as the optimization problem modeled and solved to derive the best ways to invest in these safeguards given a limited available budget. 
In Section \ref{sec:evaluation}, we undertake comparisons of the game-theoretic defending
strategies against alternative common-sense approaches as well 
as we plot the results of the Knapsack optimization to illustrate the optimal 
investment solutions. Finally, Section 5 concludes this paper by summarizing its 
main contributions and highlighting future work to be undertaken to further improve 
the performance and the usability of our model.

\section{Related Work}\label{sec:relatedwork}

This work has been inspired by a previous work of Fielder et al. \cite{fielder2016decision} 
where the authors have proposed decision support methodologies for the optimal 
choice of cybersecurity controls within an investment budget. 
They have addressed cybersecurity investment decisions by proposing different 
approaches; a game-theoretic approach, a combinatorial optimization approach 
and a mix of both called \textit{hybrid}.
This paper utilizes the latter method to recommend the optimal choice of safeguards 
for healthcare organizations. In this section, we discuss two classes of work 
relevant to this paper: literature on \textit{cyber hygiene in healthcare} - 
more specifically on the \textit{user-oriented cybersecurity safeguards}, 
and literature on \textit{optimal selection of cybersecurity safeguards}. 
Note that the literature covered on optimal selection of cybersecurity 
safeguards mainly highlight work beyond the literature covered in 
\cite{fielder2016decision}.

\subsection{Cyber Hygiene in Healthcare}

There have been growing concerns that the existing cybersecurity 
posture of healthcare organizations are insufficient and this has 
already impacted the confidentiality \cite{kruse2017cybersecurity} 
and integrity of medical data \cite{fernandez2017shared}. 
Further, many healthcare organizations are still using legacy 
systems such as Windows XP and Windows NT $3.1$ which Microsoft 
has long stopped 
supporting\footnote{\scriptsize{\url{https://www.itpro.co.uk/public-sector/27740/nine-in-10-nhs-trusts-still-use-windows-xp.}}}, allowing 
adversaries to easily breach the defenses (e.g., WannaCry attacks 
on NHS\footnote{\scriptsize{\url{https://www.nao.org.uk/wp-content/uploads/2017/10/Investigation-WannaCry- cyber-attack-and-the-NHS-Summary.pdf.}}}).
% \cite{national2017NHS}). 
In general, healthcare organizations being rich sources of valuable data and 
relatively weaker security postures have become attractive targets for 
cybercrime \cite{coventry2018cybersecurity}. The weaker security posture 
that they exhibit is primarily due to lack of adequate cybersecurity budget resulting in 
limited access to technology and expertise \cite{kotz2016privacy}. 

Besides, investment in cybersecurity has not been traditionally considered essential
for healthcare systems as emphasis has predominantly been upon providing 
patient care and people believed that there would be no motivation to 
attack them. 
% However, recent findings have illustrated that healthcare 
% data is considerably more valuable than any other data 
% \cite{george2018cyber,czeschik2018black}.      
On the other hand, the increasing use of \ac{IoT} technologies in healthcare has widened 
the attack surface beyond electronic health record databases and privacy issues to physical 
safety \cite{loukas2015cyber}. Alongside technical aspects, the role of the user in cybersecurity is paramount, as a significant proportion of attacks target the users directly through deceptive means such as application masquerading and spear-phishing. This is particularly the case in healthcare as deceiving a nurse, doctor, healthcare IT professional or administrator can 
impact the privacy and physical safety of patients \cite{billingsley2016cybersecurity}. 
% Halperin et al. \cite{halperin2008pacemakers} have demonstrated that implantable 
% medical devices (e.g., pacemakers and cardioverter defibrillator) are susceptible to 
% adversarial interference (remotely) violating not only the integrity and confidentiality 
% of patients' data and medical telemetry but also can compromise patients' physical safety.    

With the increasing usage of technology, the role that humans play in underlying 
security processes will continually expand. Heartfield and Loukas \cite{heartfield2018detecting} 
have developed a framework involving humans to effectively detect and report semantic social 
engineering attacks against them. Their results illustrate that involving users significantly 
improves the cyber threat detection rate affirming the importance of the human in cybersecurity. This further depicts that humans can no longer be seen as a threat and/or vulnerability in cybersecurity.   

Acknowledging the importance of human in cybersecurity along with the increase in the severity of breaches, security experts, policymakers and governments are urging to improve cyber hygiene. 
Such et al. \cite{such2019basic} have demonstrated that Cyber Essentials\footnote{\scriptsize{\url{https://www.gov.uk/government/publications/cyber-essentials-scheme-overview.}}}
have worked well for SMEs in mitigating threats exploiting vulnerabilities remotely using 
commodity-level exploitation tools. From a human-cyber interaction perspective, 
Vishwanath et al. \cite{vishwanath2019cyber} have demonstrated that cyber hygiene practices 
positively impact individuals' cyber attitude which is pivotal to cyber safety. 
These studies have actively exhibited that even general concepts of basic cyber hygiene work 
in different organizational contexts and can convincingly reduce cyber risk. 

Security training in healthcare has been studied for over 20 years. It ranges from an exploratory analysis of the factors that healthcare professionals need to focus on, up to highly targeted digital applications (e.g., \cite{zhou2018mobile}) and platforms for raising 
awareness of healthcare data privacy and security risks. 
Furnell et al. \cite{furnell1997addressing} discussed the necessity to promote information 
security issues and the need for appropriate training and awareness initiatives in healthcare 
institutions. They have highlighted factors to consider while designing training and awareness 
programmes to familiarize healthcare personnel with basic security concepts and procedures. 

% To achieve security training, work has been done in the field using gamification. 
% For example, Chen et al. \cite{chen2019self} propose the design of ``Hacked Time", 
% a desktop game that aims at encouraging cybersecurity behavior change. 
The effect to which security training and awareness programmes work for different users 
has been studied from multiple angles. The authors have shown that specifically for 
deception-based attacks, such as semantic social engineering \cite{heartfield2016taxonomy}, 
where self-study and work-based training are considerably more effective than formal 
education in cybersecurity \cite{heartfield2016you}. Besides, the perceived origin 
of training materials i.e., from security experts, third party agencies, or peers can 
have large impacts on security outcomes \cite{wash2018provides}.

% Here, we focus on security training and awareness programmes 
% for cyber hygiene, which is control CIS-17.

\subsection{Optimal Selection of Cybersecurity Controls}

Cybersecurity has become a key factor in determining the growth of organizations 
relying on information systems as it is not only a defensive measure but also has 
become a strategic decision providing a competitive advantage over rivalry firms. 
Further, the potential loss due to cyber incidents has encouraged organizations 
to imperatively consider cybersecurity investment decisions, especially in deriving the optimum level of investments between risk treatment options. 
The objective of cybersecurity investment methodologies is to compute an optimal 
distribution of cybersecurity budget and one of the initial work studying this was performed by Gordon and Loeb \cite{gordon2002economics}. 

Beyond previous works such as \cite{fielder2016decision,fielder2018risk,fielder2014game}
and the related work investigated there, Nagurney et al. \cite{nagurney2017supply} have proposed a game-theoretic supply chain network model with retailers competing to maximize their expected profits. This maximization is based on determining optimal product transactions and cybersecurity investments under budget constraints. Along the direction of optimal cybersecurity investments, Wang \cite{wang2019integrated} investigated the cybersecurity investment balance between acquiring knowledge and expertise, and deploying mitigation techniques. On the other hand, Chronopoulos et al. \cite{chronopoulos2017options} have opted a real options approach to analyze the performance of optimal cybersecurity controls on organizations. In particular, the authors have analyzed the effects of the cost of cyber attacks and the time of arrival of cybersecurity controls on the organization's optimal strategy. Similar to these papers, our work also considers the choice of the optimal strategy based on the efficacy of the control towards mitigating cyber risks.  

% To achieve realistic cybersecurity investment models, researchers have investigated investment models with uncertainties such as uncertainty in vulnerability assessment \cite{zhang2018decision} and uncertainty in risk assessment \cite{fielder2018risk}. They have also derived Nash Defending Strategies under these uncertainties showing that cybersecurity investment models are capable of providing effective decision support even in presence of uncertainty.   

% Paul and Wang \cite{paul2019socially} investigated the optimal balance between prevention, and detection and containment safeguards under uncertainty. The authors have presented that adjusted prevention impacts social cost and optimal configuration of safeguards the most. Further, they have identified gaps in existing cybersecurity frameworks' reliance on prevention and have proposed recommendations addressing the gaps. 
% \iMP{what is the difference with our paper?}
% In the direction of cybersecurity resilience, \cite{dutta2019cyber} have modeled cybersecurity resilience based on the needed security controls to facilitate defined security functions. Considering affordable residual risk, budget, resiliency and usability constraints, the authors have proposed an optimal selection of critical security controls for optimal and resilient risk mitigation planning.  

Most closely, in terms of methodology, related recent work on optimal cybersecurity investment is \cite{martinelli2018optimal} where the authors have investigated the balance between investing in self-protection and cyber insurance. The key difference is that their optimization minimizes expected risk and cyber insurance premium, while our model optimizes considering the efficacy of control in mitigating the aggregated residual risk and the security investment budget. Besides this, our work 
uses a unique combination of game theory and combinatorial optimization inspired by \cite{fielder2016decision}.

\section{Optimal Cyber Hygiene Safeguards Model}\label{sec:model}

\subsection{System Model}
Our model assists in acquiring an optimal selection of safeguards using game theory and combinatorial optimization. We assume $\mc{U}$ be the set of potential user groups consisting of employees of a healthcare organization. Any employee of a user group being susceptible to malicious activities can use any of the safeguards from the set of available safeguards $\mc{S}$ to improve their defense posture. However, each safeguard has a set of implementation levels $\mc{L}$ with each level having different efficacies in improving the security posture of user groups. 

Each user group $i$ is associated with an impact value which expresses the level of expected damage to the healthcare organization, given a successful attack against a user of a group $i$. This impact is equivalent to the overall asset value in association with user group $i$ and may relate to \emph{confidentiality}, \emph{integrity}, and \emph{availability}. We further consider $A_i$ to be a random variable that expresses the overall value of the assets that the user group $i$ has access to. For simplicity, we let the users of a group have the same \textit{access privileges}, thus having access to assets of the same value. Users of different groups have different \textit{access privileges} due to their different roles (e.g., IT personnel, healthcare practitioners, and administration) and access to different assets. The vulnerability of a user group $i$, i.e., the probability of being compromised by an attack, is captured by the security level $S_i$ exhibited by the user group $i$. We assume that $S_i$ increases with the number of safeguards applied as well as their application level.

Furthermore, we denote $R_i$ as the threat occurrence, i.e., the probability of a threat to attack the $i$ user group, and $L_i$ as the expected loss associated with a user group $i$. Using the well-known risk assessment formula, risk = (likelihood of being attacked) x (probability of success of this attack) x probable loss \cite{whitman2011principles}, we compute the risk as

\begin{equation}\label{eq:risk}
    L_i = R_i \, S_i \ A_i .
\end{equation}

\ifExtended
\iMP{we can breakdown the risk into C,I,A risks for better modeling}
\fi

An attack against a user group $i$ is partially mitigated by the efficacy value of the implemented cybersecurity safeguard $\plan$. The efficacy parameter, modeled as a random variable, depends on the selected application level and can be represented as $E(j,i) \colon\levels \times \users \rightarrow [0,1)$. 
It is evident from real-world practices that different implementation levels work differently on different users and this has motivated us in considering $E(j,i)$ rather than a single efficacy value for the level $j$ against all user groups $i$. Note that $E(j,i)$ is determined by the application level $j$ and the user group $i$. Due to the existence of 0-day vulnerabilities, we assume that $E(j,i) \neq 1$.

\begin{remark}
Different users have different likelihood of adopting a measure. A cyber hygiene measure works only when it is adopted, and this adaption rate distinguishes human users from systems. For example, a user may decide not to implement a cyber hygiene measure due to unfitting usability (e.g, hard to remember complex passwords) even if the optimization framework recommends otherwise. 
\end{remark}

Let $S(j,i)$ be the security level of a user group $i$ when level $j$ is implemented and can be expressed as $S(j,i)=1-E(j,i)$. Replacing $L_i$ and $S_i$ as $L(j,i)$ and $S(j,i)$, respectively, in formula \ref{eq:risk}, we compute the \textit{cybersecurity loss} for a safeguard application level $j$ and target $i$ as  
\begin{equation}\label{eq:loss}
    L(j,i) = R_i \, A_i \, [1-E(j,i)] .
\end{equation}

Equation \ref{eq:loss} implies the expected damage of the Defender when a user group $i$ is successfully compromised given the investigated safeguard has been applied at level $j$.

% Costs
While the application of a cybersecurity safeguard strengthens the defense of
the healthcare organization, it is associated with two types of cost namely; 
\emph{indirect} and \emph{direct}. Examples of indirect cost are System 
Performance and Usability. We express the indirect cost of an application level $j$ by the random variable $C\colon\mc{C} \times \levels \times \users \rightarrow \mathbb{Z^+}$. Note that $C(j,i)$ adheres to the defined property for any safeguard against a user group $i$. Further, the indirect cost increases with an increase in the level of application of the safeguard i.e., 

\begin{equation}
	j > j' \Leftrightarrow C(j,i) \geq C(j',i),\quad \forall j \neq j' .
\end{equation}

From the above, we derive the \textit{overall expected loss} of the 
organization when application level $j$ is applied on user group $i$ as

\begin{equation}
  \sum_{i=1}^{|\users|} L(j,i) + C(j,i) .
\end{equation}

Each level has also a direct cost expressed by the random variable $F\colon \levels \rightarrow \mathbb{Z^+}$ that maps the safeguards and application levels to the monetary cost of the plan. In this paper, we refer the direct cost to be the available investment budget of the organization. For reference purposes, the symbols used throughout this paper are described in Table \ref{tab:list_symbols}.

\begin{table}[t]
\footnotesize
\centering
\renewcommand*{\arraystretch}{1.1}
\begin{tabular}{c l}
  \hline
  \textbf{Symbol} & \textbf{Description} \\ 
  \hline
  $\safeguards$    			& Set of safeguards \\ 
  $\users$    			    & Set of users \\ 
  $\levels$                 & Set of safeguard implementation levels \\
  $R_i$                     & Probability of group $i$ to be attacked\\
  $S_i$                     & Security level of group $i$\\
  $A_i$                     & Asset value that group $i$ has access to\\
  $\lambda$                 & Maximum application level\\
  $U_d$                     & Utility of the Defender \\
  $U_a$                     & Utility of the Attacker \\
%   $\mixdef$                 & Randomized Safeguard Strategy \\
%   $\mixdef(j)$              & Probability of applying level $j$ \\
%   
$\mixdef_{\sigma,j}$      & Randomized Safeguard Strategy for safeguard $\sigma$ at application level $j$\\
  $\mixatt$                 & Randomized Attacking Strategy \\
  $\mixatt(i)$              & Probability of attacking group $i$ \\
  $L_i$                     & Expected loss from group $i$ \\
  $L(j,i)$                  & Expected loss from group $i$ when choosing application level $j$\\
  $L(\mixdef_{\sigma,j},i)$    & Expected loss from group $i$ when choosing Safeguards Plan $\mixdef_{\sigma,j}$\\
  $C(j,i)$                  & Indirect cost of level $j$ when applied to group $i$\\
  $E(j,i)$                  & Efficacy of application level $j$ on group $i$ \\
  $E(\mixdef_{\sigma,j},i)$    & Efficacy of safeguards plan $\mixdef_{\sigma,j}$ on group $i$\\
  $\game$                   & Cyber Safeguard Game for safeguard $\sigma$ and maximum application level $\lambda$\\
  $\equilibrium$            & Nash Safeguards Plan\\
  $F(\mixdef_{\sigma,j})$      & Financial cost of Safeguards Plan $\mixdef_{\sigma,j}$\\
  $F(\sigma,j)$             & Financial cost of safeguard $\sigma$ when applied at level $j$\\  
  $B$                       & Available financial budget to invest in Nash Safeguards Plans\\
  \hline
\end{tabular}
\vspace{0.3cm}
\caption{List of Symbols}
\label{tab:list_symbols}
\vspace{-0.8cm}
\end{table}

\subsection{Game-Theoretic Model for Selection of Safeguards Levels}

This section presents a formal model for the selection of safeguard implementation levels for each of the available safeguards. The Defender chooses to implement (or apply as in this paper we use these two terms interchangeably) 
a cyber hygiene safeguard from $\safeguards$, while the Attacker chooses to attack a user group from $\users$. The Defender must decide to apply this safeguard at a specific level (pure strategy) or combination of different levels (mixed strategy) both from $\levels$. The higher the level, the greater is the applied degree of a cyber hygiene safeguard. We refer to the application of a safeguard $s$ at a certain level $j$ as \textit{cybersecurity safeguard plan}. This strategic interaction is modeled as a game where the Defender chooses the level of a safeguard to implement rather than the safeguards from $\safeguards$.
% We investigate the optimal strategic decisions of Defender
% regarding the application levels for each safeguard.

We define the Cyber Safeguard Game (CSG) between Defender and Attacker, as an \emph{one-shot, bimatrix} game of \emph{complete information} played for any of the safeguards leading to a total number of $\mc{S}$ independent games. For simplicity, we have assumed no inter-dependencies between the safeguards, i.e., each safeguard mitigates a portion of the overall risk inflicted by the Attacker \cite{smeraldi2014spend}.

The set of pure strategies of the Defender consists of all possible application levels, $j \in \levels$, while the Attacker's pure strategies are the different user groups $i \in \users$ which could be targeted using attacks such as social engineering. Thus, in CSG a pure strategy profile is a pair of Defender and Attacker actions, $(j,i) \in \levels \times \users$ giving a pure strategy space of size $|\levels| \times |\users|$. For the rest of the paper, we adopt the convention where the Defender is the row player and the Attacker is the column player.
% Furthermore, since in each CSG we investigate \textit{one and only one} safeguard, we
% abuse notation by referring to efficacy of the implementation level $j$ 
% on a target $i$ by $E(j,i)$, the indirect cost by $C(j)$, and the 
% associated cybersecurity loss by $L(j,i)$.
% The latter has been previously defined by \ref{eq:loss}.

Each player's preferences are specified by a \emph{payoff function} defined as $U_d:(j,i) \rightarrow \mathbb{R_{-}}$ and $U_a:(j,i) \rightarrow \mathbb{R_{+}}$ for the Defender and Attacker, respectively, for the pure strategy profile 
$(j,i)$. According to \cite{osborne1994course}, we define a \emph{preference relation} $\succsim$, when $i$ is chosen by the Attacker, defined by $j \succsim j'$, if and only if $U_d(j,i) \geq U_d(j',i)$. In general, given the set $\levels$ of all available application levels of a safeguard, a rational Defender can choose a level (i.e., pure strategy) $j^*$ that is \emph{feasible}, that is $j^* \in \levels$, and \emph{optimal} in the sense that $j^* \succsim j, \forall j \in \levels, j\neq j^*$; alternatively she solves the problem $\max_{j \in \levels} U_d(l, i)$, for a user group $i \in \users$. Likewise, we define the preference relation for the Attacker, where $i \succsim i' \iff U_a(j,i)\geq U_a(j,i')$, for an application level $j \in \levels$. CSG is a game defined for each cyber hygiene safeguard and it is realistic to assume that all levels may be available for selection by the Defender. Their availability depends on the investment budget of the Defender and the overall financial cost of the game solution. 

To derive optimal strategies for the Defender, we deploy the notion of 
\emph{mixed strategies}. Since players act independently, we can enlarge 
their strategy spaces to allow them to base their decisions on the outcome 
of random events that create uncertainty to the opponent about individual 
strategic choices maximizing their payoffs. Hence, both Defender and Attacker 
deploy randomized (i.e., mixed) strategies. The mixed strategy $\mixdef$ 
of the Defender is a probability distribution over the different application 
levels (i.e.~pure strategies) where $\mixdef(j)$ is the probability of 
applying level $j$ under mixed strategy $\mixdef$. We refer to a mixed strategy of the Defender as a \emph{Randomized Safeguard Strategy} (RSS). For the finite nonempty set $\levels$, let $\Pi_{\levels}$ represent the set of all probability distributions over it, i.e.,
\begin{eqnarray}\label{eq:set_probs_L}
    \Pi_{\levels} := \{\mixdef \in 
    \mathbb{R}^{+R} |~\sum_{j \in \levels} \mixdef(j)=1 \} .
\end{eqnarray}

Therefore a member of $\Pi_{\levels}$ is a mixed strategy of the Defender.
Likewise, the Attacker's mixed strategy is a probability distribution over the 
different available user groups. This is denoted by $\mixatt$, where 
$\mixatt(i)$ is the probability of attacking the $i$-th user
group under mixed strategy $\mixatt$. We refer to a mixed strategy of the Attacker as the \emph{Randomized Attacking Strategy} (RAS). Alike (\ref{eq:set_probs_L}),
we express $\Pi_{\users}$ as the set of all probability distributions over the set of all Attacker's pure strategies (i.e., given by $\users$). Therefore, a member of $\Pi_{\users}$ is as a mixed strategy of the Attacker. From the above, the set of mixed strategy profiles of CSG is the Cartesian product of the individual mixed strategy sets, $\Pi_{\levels} \times \Pi_{\users}$.	

\begin{definition}(Support of RSS)
    The support of $\mixdef$ is the set of application levels 
    $\{j|\mixdef(j)>0\}$, and it is denoted by $supp(\mixdef)$.
\end{definition}

\begin{definition}(Support of RAS)
    The support of $\mixatt$ is the set of healthcare user groups 
    $\{i|\mixatt(i)>0\}$,~and it is denoted by $supp(\mixatt)$.
\end{definition}

The above definitions state that the subset of applications levels 
(resp. user groups) that are assigned positive probability by the mixed 
strategy $\mixdef$ (resp. $\mixatt$) is called the \emph{support} of 
$\mixdef$ (resp. $\mixatt)$). Note that a pure strategy is a special 
case of a mixed strategy, in which the support is a single action.

Now that we have defined the mixed strategies of the players, we define 
CSG as the finite strategic game 
\begin{equation}
    \Gamma:=\langle(\mathrm{Defender},~\mathrm{Attacker}),\Pi_{\levels} 
    \times \Pi_{\users},~(U_d,U_a)\rangle .
\end{equation}
For a given mixed strategy profile $(\mixdef,\mixatt) \in \Pi_{\levels} \times \Pi_{\users}$, we denote by $U_d(\mixdef,\mixatt)$, and $U_a(\mixdef,\mixatt)$ the expected payoff values of the Defender and Attacker, where the expectation is due to the independent randomization according to mixed strategies $\mixdef$,~and $\mixatt$. This can be formally represented as  

\begin{equation}\label{eq:util_def}
    \begin{aligned}
    U_d(\mixdef,\mixatt):=\sum_{j \in \levels} \sum_{i\in \users} U_d(j,i)  
    \,\mixdef(j) \,\mixatt(i) , 
    \end{aligned}	
\end{equation}

and similarly 
\begin{equation}\label{eq:util_att}
    \begin{aligned}
    U_a(\mixdef,\mixatt):=\sum_{j\in \levels} \sum_{i\in \users} 
    U_a(j,i) \,\mixdef(j) \,\mixatt(i).
    \end{aligned}
\end{equation}

By using the preference relation we can say that, for an Attacker's mixed strategy $\mixatt$, the Defender prefers to follow the RSS $\mixdef$ 
as opposed to $\mixdef'$ (i.e., $\mixdef \succsim \mixdef'$), 
if and only if $U_d\mixed \geq U_d(\mixdef',\mixatt)$.

\begin{definition}
    The Defender's (resp. Attacker's) best response to the mixed strategy 
    $\mixatt$ (resp. $\mixdef$) of the Attacker (resp. Defender) is an RSS 
    (resp. RAS)
    $\mixdef^{\BR} \in \Pi_{\levels}$ (resp. $\mixatt^{\BR} \in \Pi_{\users})$ 
    such that $U_d(\mixdef^{\BR},\mixatt) \geq U_d(\mixdef,\mixatt), 
    \forall \mixdef\in \Pi_{\levels}$ (resp. $U_a(\mixdef,\mixatt^{\BR}) 
    \geq U_d(\mixdef,\mixatt), \forall \mixatt\in \Pi_{\users})$.
\end{definition}

\begin{remark}
The game-theoretic solutions that we propose in the next section involve \emph{randomization}. For instance, in a mixed equilibrium, each player's randomization leaves the other \emph{indifferent} across her randomization support. These choices can be deliberately randomized, however these are not the 
only equilibria interpretations. For instance, the probabilities over the pure actions (i.e., application level or user group pure selections) can represent (i) time averages of an ``adaptive'' player, (ii) a vector of fractions of a ``population'', where each player type adopts pure strategies and, (iii) a ``belief'' vector that each player has about the other regarding their behavior. 
\end{remark}

%
% CSG solutions
%
\subsection{CSG solutions}\label{solutions}

Given the definition of CSG and its components, we derive optimal strategies for the Defender. First, we investigate the problem of determining best RSSs and RASs (i.e., mixed strategies), for the Defender and the Attacker respectively, when both players are strategic and play simultaneously. 
% A \emph{game solution} is a prediction of how rational players may take decisions.

As we have not explicitly defined the \emph{strategic type} of Attacker, 
we consider different types of solutions based on various Attacker behaviors.
This analysis will allow us to draw robust conclusions regarding the 
\emph{overall optimal} Defender strategy, which will minimize expected damages 
\emph{regardless of the Attacker type}.

% \subsection{Nash mixed strategies}
The most commonly used solution concept in game theory is that of 
\emph{Nash Equilibrium} (NE) \cite{osborne1994course}. This concept 
captures a steady state of the play of the CSG in which both Defender and 
Attacker hold the correct expectation about the other players' behavior 
and they act rationally. A NE dictates optimal responses to each other's 
actions, keeping the others' strategies fixed, i.e., strategy profiles that 
are resistant against unilateral deviations of players.

\begin{definition}
    In any Cyber Safeguard Game, a mixed strategy profile 
    $(\mixedNE)$ of $\Gamma$ is a mixed NE if and only if 
    \begin{enumerate}
    	\item $\mixdef^{\NE} \succsim \mixdef, \forall \mixdef \in \Pi_{\levels}$,
    	when the Attacker chooses $\mixatt^{\NE}$, i.e.
    	\begin{eqnarray}
    		U_d(\mixedNE)\geq_{\forall \mixdef\in \Pi_{\levels}} U_d(\mixdef,\mixatt^{\NE});
    	\end{eqnarray} 
    	\item $\mixatt^{\NE}\succsim \mixatt, \forall \mixatt \in \Pi_{\users}$, 
    	when the Defender chooses $\mixdef^{\NE}$, i.e.
    	\begin{eqnarray}
    		U_a(\mixedNE)\geq_{\forall \mixatt\in \Pi_{\users}} U_a(\mixdef^{\NE},\mixatt).
    	\end{eqnarray} 
    \end{enumerate}
\end{definition}

\begin{definition} 
    The Nash Safeguards Plan (NSP),~denoted by $\mixdef^{\NE}$, is a probability 
    distribution over the different levels, as determined by the NE of the CSG.
\end{definition} 
% \iMP{I refer to equilibria of games as Plans, and to the Knapsack solutions as Investment Strategies}

\textbf{Example 1}. For a safeguard with 3 application levels including level 0, which 
corresponds to not applying the safeguard at all, an NSP $(0,0.2,0.8)$ dictates 
that 20\% of the users will be strengthened (e.g., trained) at $j=1$ (e.g., 
once when they join the organization), while 80\% of the users will be applied 
a higher level of the safeguard $j=2$ (e.g., attending training once per year).
% Note that this distribution does not determine a unique  
% application level to be chosen, as this decision is probabilistic.

%
% Optimality analysis
%
\subsection{Optimality analysis}\label{analysis}

We model \emph{complete information} Nash CSGs, according to which both 
players know the game matrix, which contains the utilities of both 
players for each pure strategy profile. The utility function of the 
Defender is determined by the probability of failing to protect a user group and the indirect costs of the chosen application levels.
We consider a \emph{zero-sum} CSG, where the Attacker's utility is the 
opposite of the Defender's utility. The rationale behind the zero-sum CSG is that when the Defender is uncertain about the Attacker type, she considers the \emph{worst case scenario}, which can be formulated by a zero-sum game where the Attacker can cause her \emph{maximum damage}.
% \iMP{there is a survey on game-theoretic approaches for cybersecurity that explains this
% as the best choice overall}
The idea behind a zero-sum game like this is that the Attacker focuses on causing
maximum corruption to cyberspace, while the Defender aims at minimizing the damage. Due to the Attacker's goal being conflicting to the Defender's objective, the application of game theory to study the selection of safeguards application levels is convenient. While in most security situations the interests of the players are neither in strong conflict nor in complete identity, the zero-sum game provides important insights into the notion of ``optimal play'', which is closely related to the \emph{minimax theorem} \cite{minimax}.

In the zero-sum CSG, 
\begin{equation}
    \zs=\langle \{d,a\}, \levels \times \users, \{U_d,-U_d\}\rangle , 
\end{equation}
the Attacker's gain is equal to the Defender's security loss, and vice versa.
We define the utility of the Defender in $\zs$ as 
\begin{eqnarray}\label{eq:utility_defender_in_zs}
 \small
 U_d^{\zs}(j,i) := - w_L \, L(j,i) - w_C \, C(j,i).
\end{eqnarray}
The first term of (\ref{eq:utility_defender_in_zs}) is the expected  
loss of the Defender inflicted by the Attacker when 
attempting to compromise user group $i$, while the second term 
expresses the aggregated indirect cost of the safeguard application 
irrespective of the attacking strategy.
Let $w_L, w_C \in [0,1]$ are importance weights, which can facilitate 
the Defender with setting her preferences in terms of security loss, 
and indirect cost, accordingly. 
% \iMP{we can vary these for more results during evaluation}

For a mixed profile $(\mixdef,\mixatt)$, the utility of the Defender equals
\begin{equation}
\label{eq:mixed_payoff_def}
    \small
    \begin{aligned}
        U_d^{\Gamma_0}(\mixdef,\mixatt) &\overset{(\ref{eq:util_def})}{=}
        \sum_{j \in \levels} \sum_{i \in \users} U_d^{\Gamma_0}(j,i)  \mixdef(j) \, \mixatt(i)\\ 
        &\overset{(\ref{eq:utility_defender_in_zs})}{=}
        \sum_{j\in \levels} \sum_{i\in \users} [-w_L \, L(j,i) - w_c \, C(j)]\,\mixdef(j) \,\mixatt(i) \\
        &= - w_L \sum_{j\in \levels} \sum_{i\in \users} L(j,i)\,\mixdef(j) \,\mixatt(i) 
        - w_C \sum_{j\in \levels} C(j,i)\,\mixdef(j) .
    \end{aligned}
\end{equation}
As $\zs$ is a zero-sum game, the Attacker's utility 
is given by $U_a^{\zs}(\mixdef, \mixatt) 
= -\,U_d^{\zs}(\mixdef, \mixatt)$. Since the Defender's 
equilibrium strategies maximize her  utility, given that 
the Attacker maximizes her own utility, we will refer to 
them as \emph{optimal strategies}.

As $\zs$ is a two-person zero-sum game with a finite number of 
actions for both players, according to Nash \cite{nash1950equilibrium}, 
it admits at least a NE in mixed strategies and saddle-points 
correspond to Nash equilibria as discussed in \cite{alpcan2010network}
(p.\,42).~The following result from \cite{basar1995dynamic}, 
establishes the existence of a saddle (equilibrium) solution 
in the games, we examine and summarizes their properties.

\begin{definition}[Saddle point of the CSG] 
    The $\zs$ Cyber Safeguard Game (CSG) admits a saddle point in mixed 
    strategies, $(\mixdef^{\NE}_{\zs},\mixatt^{\NE}_{\zs})$, with the 
    property that
    \begin{itemize} 
    \item $\mixdef^{\NE}_{\zs}=\arg \max_{\mixdef \in \Delta_{\levels}} \min_{\mixatt \in \Delta_{\users}} U_d^{\Gamma_0}(\mixdef,\mixatt),\; \forall \mixatt$, and
    \item $\mixatt^{\NE}_{\zs}=\arg \max_{\mixatt \in \Delta_{\users}} \min_{\mixdef \in \Delta_{\levels}} U_a^{\Gamma_0}(\mixdef,\mixatt),\,\forall \mixdef$.
    \end{itemize}
    Then, due to the zero-sum nature of the game, the minimax theorem \cite{minimax} holds, i.e.~
    $\max_{\mixdef \in \Delta_{\levels}} \min_{\mixatt \in 
    \Delta_{\users}} U_d^{\Gamma_0}(\mixdef,\mixatt)= \min_{\mixatt \in \Delta_{\users}} 
    \max_{\mixdef \in \Delta_{\levels}} U_d^{\Gamma_0}(\mixdef,\mixatt)$ .
    
    The pair of saddle point strategies $(\mixdef^{\NE}_{\zs},\mixatt^{\NE}_{\zs})$ are at the same time security strategies for the players, i.e.~they ensure a minimum performance regardless of the actions of the other.~Furthermore, if the game admits multiple saddle points (and strategies), they have the ordered interchangeability property, i.e.~the player achieves the same performance level independent from the other player's choice of saddle point strategy.
\end{definition}

The minimax theorem \cite{minimax} states that for zero-sum games, NE and minimax 
solutions coincide.~Therefore, $\mixdef^{\NE}_{\zs} = 
{\tt \arg\min}_{\mixdef \in \Delta_{\levels}} \max_{\mixatt \in 
\Delta_{\users}} U_a^{\Gamma_0} (\mixdef,\mixatt)$.
This means that regardless of the strategy the Attacker 
chooses, NSP is the Defender's security strategy 
that guarantees a minimum performance.

% We can convert $\zs$ into a Linear Programming (LP) problem and make use of some of the powerful algorithms available for LP to derive the equilibrium.~For a given mixed strategy $\mixdef$ of the Defender, we assume that the Attacker can cause maximum damage to $\destination$ by injecting a message $\widehat{m}$ into the cluster network.

Formally, the Defender seeks to solve the following LP: 
\begin{equation}
    \small
    \begin{aligned}
        &\max_{\mixdef \in \Delta_{\levels}} \min_{\mixatt \in \Delta_{\users}} U_d^{\Gamma_0}(\mixdef, \hat{i}\,) \\ 
        \text{subject} & \text{~to}
        \begin{cases} 
        \label{eq:lp}
        U_{d}^{\Gamma_0}(\mixdef,1) - \min_{\mixatt \in \Delta_{\users}}U_d^{\Gamma_0}(\mixdef, \hat{i})e \geq 0\\ \hspace{2cm} \vdots \\ U_{d}^{\Gamma_0}(\mixdef,|\users|) - \min_{\mixatt \in \Delta_{\users}}U_d^{\Gamma_0}(\mixdef, \hat{i})e \geq 0\ \\ \mixdef e = 1 \\ \mixdef \geq 0 .
        \end{cases} 
    \end{aligned}
\end{equation}
In this problem, $e$ is a vector of ones of size $|\users|$.
% For safeguard $x$, the defending strategy is a probability 
% distribution $\mixdef_x=[\mixdef_{x,1},...,\mixdef_{x,|\levels|}]$ 
% over the different choices of application levels for this safeguard.
% The different user groups are the targets for Attacker whose
% attacking strategy in a CSG is a probability distribution 
% $\mixatt=[\mixatt_{1},...,\mixatt_{|\users|}]$ over the different $|\users|$ user groups.

\subsection{Multiple Games Per Safeguard}

Given that we have to allocate a budget in applying different 
safeguards, we may come across the challenge of not having enough 
monetary resources to select some of the equilibria of the CSG. 
Therefore, one has to derive the financial cost of equilibrium 
and assess its feasibility by comparing its financial cost to the 
available remaining budget. We refer to ``remaining'' budget as
we expect that the Defender will have to select among a number of 
equilibria, one per safeguard, as we show later in this section. 

To provide to the Defender a wider variety, in terms of financial cost,
of equilibria per safeguard, we define a number of CSGs per safeguard. Each of these games has a different number of application levels available to the Defender. 
Aligned with \cite{fielder2016decision}, for each safeguard $\sigma$, we 
study $|\levels|$ CSGs. 

\begin{definition}
To differentiate among different safeguards and implementations levels, 
we denote the CSG by $\game$, where the 
safeguard $\sigma$ can be applied up to $\lambda \in [0,|\levels|]$. 
% We refer to each of the games as $\lambda$-CSG. 
\end{definition}

\begin{figure*}[t]
	\centering
	\includegraphics[width=1\linewidth]{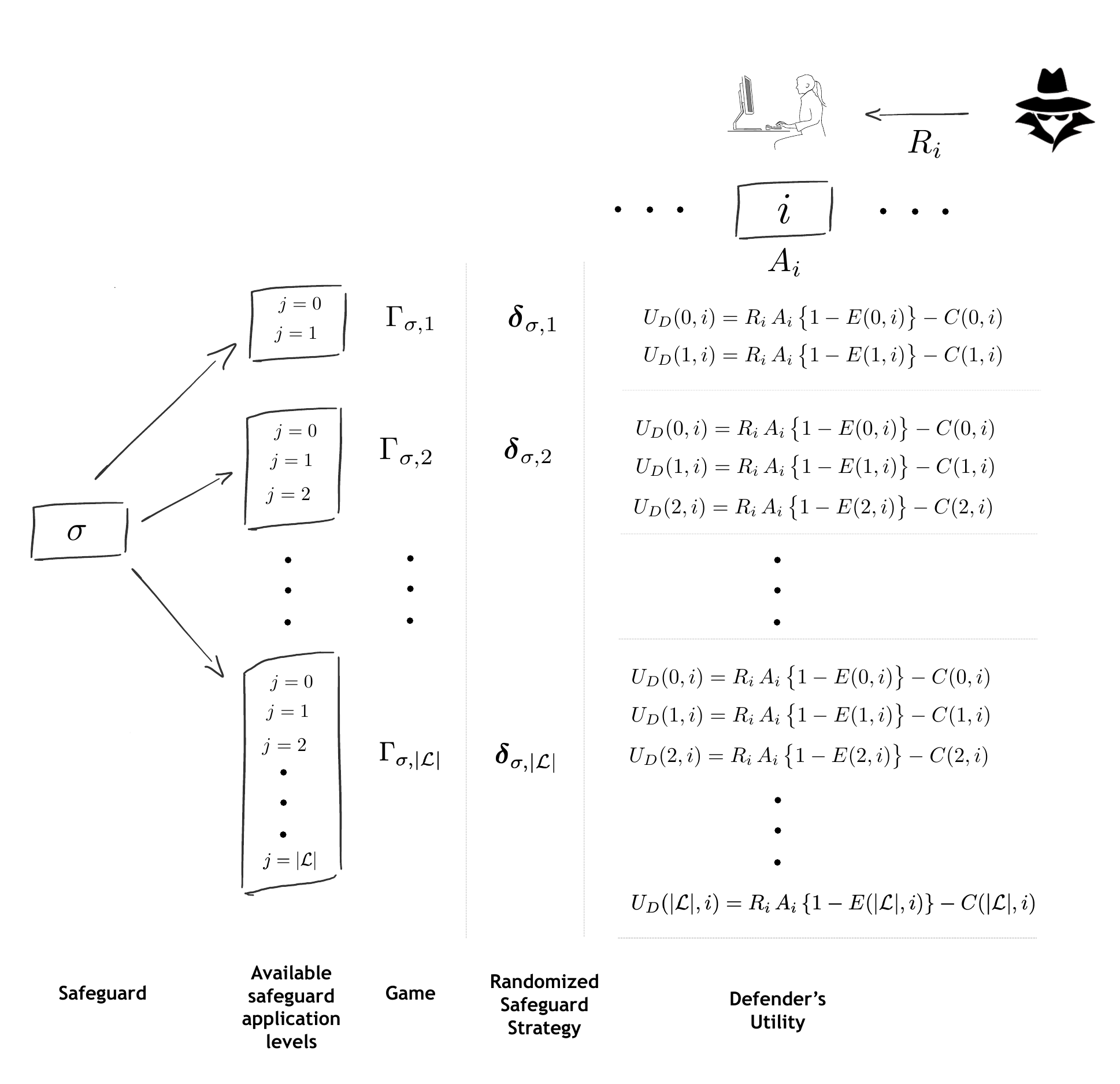}
	\caption{Illustration of the safeguard-centered model of OST used to devise game-theoretic strategies for the Defender.}
	\label{fig:gt_illustration}
\end{figure*}

Note that we allow $\lambda=0$ so that the Defender has the option to avoid selecting
a safeguard should this violate some budget constraints. 
A Knapsack optimisation is used in the second phase of the model to select the equilibria, at most one per safeguard. In this way, we manage to have $|\levels|$ NSPs per safeguard, 
each of a different financial cost. Each $\game$ is a game where 
(i) Defender's pure strategies correspond to consecutive application levels 
of safeguard $\sigma$ starting always from 0 and including all levels up to $\lambda$ 
and, (ii) Attacker's pure strategies are the different targets akin to user groups. 
Figure \ref{fig:gt_illustration} illustrates the different Cybersecurity Safeguards Games
along with the utilities of the Defender. 

Let $\equilibrium$ be the equilibrium of $\game$ then
\begin{equation}
    \equilibrium = [\delta^{NE}_{\sigma,0}, 
    \delta^{NE}_{\sigma,1}, \dots, 
    \delta^{NE}_{\pair}] .
\end{equation}

Let $F(\mixdef_{\pair})$ be the financial cost of the safeguards plan 
$\mixdef_{\pair}$ which can be derived by summing the financial costs 
of all application levels $j \in \{1,2,\dots,\lambda\}$ for safeguard $\sigma$ contributed 
proportionally by using the corresponding probability from 
$\mixdef_{\pair}$, i.e., $\delta_{\sigma,j}$.
Let $F(\sigma,j)$ denote the financial cost of safeguard $\sigma$ then

\begin{equation}
    F(\mixdef_{\pair}) = \sum_{j \in \{1,2,\dots,\lambda\}} \delta_{\sigma,j} \, F(\sigma,j) .
\end{equation}

% investing in cyber security one might face the challenge of not having a necessary financial budget to implement the equilibria of a cyber security safeguard Game. To tackle this chal- lenge we define cyber security safeguard Subgames, which constitute a safeguard Game by gradually increasing the available implementation levels of the safeguard. In this way, we can derive a number of equi- libria that can satisfy a wider range of financial capacity. A safeguard Subgame Gjk is a game where (i) D’s pure strategies correspond to consecutive implementation levels of the safeguard j starting always from 0 (i.e., fictitious safeguard-game) and including all levels up to k and, (ii) A’s pure strategies are the different targets akin to pairs of vulnerabilities and depths.

\subsection{Investment in Nash Safeguards Plans} \label{sec:investment_NSP}

Let $\safeguards$ be the set of all available safeguards 
to the Defender. We can solve all $|\safeguards| \times |\levels|$
CSGs and derive a set of equilibria per safeguard $\sigma$ represented as follows
\begin{equation}
 \{\mixdef^{NE}_{\sigma, 1},\mixdef^{NE}_{\sigma, 2}, \dots, \mixdef^{NE}_{\sigma, |\levels|}\} .
\end{equation}

For all safeguards $\{1,2,\dots,|\safeguards|\}$ the following set of sets of equilibria, i.e., NSPs, is available
\begin{equation}\label{eq:allNSPs}
 \Big\{\{\mixdef^{NE}_{1,0},\mixdef^{NE}_{1,1}, \dots, \mixdef^{NE}_{1,|\levels|}\}, 
 \{\mixdef^{NE}_{2,0},\mixdef^{NE}_{2,1}, \dots, \mixdef^{NE}_{2,|\levels|}\}, \dots,
 \{\mixdef^{NE}_{|\safeguards|,0},\mixdef^{NE}_{|\safeguards|,1}, \dots, \mixdef^{NE}_{|\safeguards|,|\levels|}\}\Big\} .
\end{equation}

Optimal budget allocation in cybersecurity can be tackled by combinatorial optimization as 
previously investigated by Smeraldi and Malacaria \cite{smeraldi2014spend}.
We are concerned with the challenge of protecting multiple targets, in our case user groups, 
with the use of a number of NSPs that interact between them in different ways. 
In the following, we model the challenge of investing in these different NSPs in a way that
at most one NSP per safeguard is chosen and the sum of financial costs of these NSPs fits an available cybersecurity budget. We have used 0-1 Knapsack Optimization to solve this
problem. As opposed to the solution provided in \cite{fielder2016decision}, we have chosen
the objective function of the Defender to consider the sum of expected losses incurred from
the different user groups being attacked. This is not to say that the proposed weakest-link 
model in \cite{fielder2016decision} is not relevant anymore but we realize the potential risk 
to have all user groups targeted by the Attacker with the goal to maximize the collective damage over 
a number of assets rather than trying to compromise the most precious asset. We argue that such
a goal to maximize the aggregated damage is more applicable in attacks like Advanced Persistent
Threat, where the goal is to maximize the Defender's overall loss in a number of different
ways.
% Knapsack mathematical optimisation. 
% We assume that Defender possesses a security budget to invest in cyber 
% hygiene safeguards that can be implemented at different levels. Each application
% level for a safeguard generates a \textit{cybersecurity hygiene process} and it comes 
% with some \textit{indirect costs}, e.g., how much reduces the productivity of users
% who adopt this process. For example, Renaud's work on a UK hospital, where she has observed that 
% ``unrealistic (security) task demands'' placed on employees had significant negative 
% consequences for both security and productive tasks \hl{[R12]}. 

% Knapsack
The Knapsack Problem (KP) is an NP-hard problem \cite{pisinger2005hard}. 
There are several applications of KP such as resource distribution, 
investment decision making and budget controlling.
In our model, we define KP as: 
Assuming that there is a knapsack with a maximum capacity of $B$, which represents the budget of the Defender.
Given the set of all possible $|\safeguards| \times |\levels|$ NSPs shown in (\ref{eq:allNSPs}), 
each Knapsack candidate solution consists of at most $|\safeguards|$ NSPs, one per each safeguard. 
Each NSP reduces, to some degree, the overall cyber risk of the organization as a result of reducing 
the individual risk on each user group. 
The problem is to select a subset of NSPs that maximize the knapsack profit without exceeding the 
maximum capacity of the knapsack. 
We define an optimal solution to our KP as 
$\Psi = \{\mixdef^{NE}_{\pair}\},\forall\sigma\in\safeguards,\forall\lambda\in\levels$.
A solution $\Psi$ takes exactly one solution (i.e., equilibrium or cybersecurity plan) for each
safeguard as a \textit{policy for implementation/application}. To represent the cyber security
investment problem, we need to expand the definitions for both expected loss $L$ and effectiveness $E$ 
to incorporate the solutions of the different CSGs. Hence, we expand $L$ such that 
$L(\mixdef_{\pair},i)$ is the expected loss inflicted by compromising user group $i$ given the 
application of the plan $\mixdef_{\pair}$.
We also expand $E$ such that 
$E(\mixdef_{\pair},i)$ is the efficacy that $\mixdef_{\pair}$ brings when applied to 
user group $i$.
From equation (\ref{eq:loss}) the expected loss on user group $i$ when 
NSP $\mixdef_{\pair}$ is applied is given by
\begin{equation}\label{eq:loss_plan}
    L(\mixdef_{\pair},i) = R_i \, A_i \, [1-E(\mixdef_{\pair},i)]
\end{equation}
A natural approach is the KP to seek a set of NSPs that minimize the aggregated expected risks 
across all user groups. We assume that each NSP may protect more than one user groups. 
We then seek optimal safeguards allocation for a series of user groups each of which 
can be protected by a different set of NSPs. The latter may not necessarily have an 
additive efficacy. The following illustrated example considers two NSPs and explains
how we have decided to combine their efficacy in a single formula that we then use in KP 
formulation. 

\textbf{Example 2}. 
By slightly abusing notation, assume two 
NSPs $\mixdef, \mixdef'$ that mitigate 20\% and 30\% of the same 
user group risk, respectively.
If the NSPs had additive efficacy the total expected loss 
on user group $i$ when applying both $\mixdef, \mixdef'$
equals
$R_i \, A_i \, \Big\{1 -\big\{E(\mixdef,i) + E(\mixdef',i)\big\}\Big\} = 
R_i \, A_i \, (1-0.2-0.3) = 0.5 \, R_i \, A_i$.
In this paper, we assume a more conservative expected loss mitigation 
function when combining two or more NSPs as follows
$R_i \, A_i \, \Big\{\big\{1-E(\mixdef,i)\big\} \, 
\big\{1-E(\mixdef',i)\big\}\Big\} =
R_i \, A_i \cdot (1-0.2) \, (1-0.3) =
R_i \, A_i \cdot 0.8 \cdot 0.7 = 
0.56 \, R_i \, A_i$. 
% It is worth noting that in the recent US patent \cite{granadillo2019selection}, 
% authors demonstrate the union and intersection of countermeasures with the use of a 3D shape. 

Given the above, if we represent the solution $\Psi$ by the bitvector $\vec{z}$, 
we can then represent the 0-1 KP as
\begin{eqnarray}\label{eq:knapsack}
    && \max_{\vec{z}} \sum_{i=0}^{|\users|} A_i \, R_i  
    \Bigg\{\prod_{\sigma=1}^{|\safeguards|} 
    \Big\{1 - \sum_{j=0}^{\lambda} E(\mixdef_{\pair}^{NE},i) \, z_{\pair} \Big\}\Bigg\} \nonumber \\
	&& \text{s.t.}~\sum_{\sigma=1}^{|\users|} \sum_{\lambda=0}^{|\levels|} 
	F(\mixdef_{\pair}) \, z_{\pair} \leq B, \nonumber \\
	&& \sum_{\lambda=0}^{|\levels|} z_{\pair}=1, z_{\pair}\in\{0,1\}, \forall \sigma=1,2,\dots,|\safeguards| .
\end{eqnarray}
where $B$ is the available budget of the Defender to be spent in cyber safeguards and 
$z_{\pair}=1$ holds when $\mixdef_{\pair}^{NE} \in \Psi$. Among KP solutions that all 
maximize the overall expected loss, we choose the solution with the lowest financial cost as this will be, in overall, the best advice to the defender producing same benefit for lower price.

%
% If Extended
%
\ifExtended
However, in general such a budget allocation is not a given and should be a 
product of the optimisation. We can obtain the budget allocation by 
noticing that the dynamic programming algorithm above actually finds an 
optimal allocation of resources for all budgets up to B.
\fi

\section{Model Evaluation}\label{sec:evaluation}

We have developed the proposed models
as part of the Optimal Safeguards Tool (OST) proposed in \cite{mohammadi2019curex}. 
OST computes Nash Safeguards Plans as well as the Knapsack solutions.
OST aims at offering realistic actionable advice to healthcare organizations. 
The following represents a case study based on Critical Internet Security (CIS) 
17 Control ``Implement a Security Awareness and Training Program''.

\subsection{Use Case}

\subsubsection{User groups.}
Here, we assume a representative (non-exhaustive) set of three user groups, 
denoted by $i$, in decreasing order of \textit{access privileges}: 
\begin{itemize}
    \item $i=1$; \textbf{ICT}: The information and communication technology 
    professionals responsible for the systems, networks and software. 
    They set up digital systems, support staff who use them, diagnose and 
    address faults, as well as set up and maintain security provisions. 
    In addition to the ICT infrastructure, they may also interact with medical 
    devices and electronic healthcare record systems. We consider the value of 
    corresponding assets that can be affected by an attack on this group to be 
    the highest possible, $A_1=100$ (e.g., \$100k). At the same time, due to limited interaction 
    with the public, this is the group with the lowest visibility to attacks 
    targeting the human, and as such we can consider it as lower risk, $R_1=0.2$.
    
    \item $i=2$; \textbf{Clinical}: Nurses, doctors and other clinical staff have 
    access to medical devices and electronic healthcare records. We consider the 
    value of corresponding assets that can be affected by an attack on this group 
    to be $A_2=50$ (e.g., \$50k). As a result of visibility due to interaction with the patients 
    and presence on the hospital's website, this group has a moderate risk, $R_2=0.5$.
    
    \item $i=3$; \textbf{Administration}: Receptionists, medical secretaries and other 
    administration roles involve access to electronic healthcare records. We consider 
    the value of corresponding assets that can be affected by an attack on this group 
    to be $A_3=25$ (e.g., \$25k). This group of users may have high interaction with the public 
    and volume of email traffic (e.g., appointment requests) and as such high risk, 
    $R_3=0.8$.
\end{itemize}

\begin{table}[th]
    \begin{center}
    \label{tab:cis174}
    \begin{tabular}{@{}lcccc@{}}
    \toprule
    Control level \textbackslash Role                             &                        & ICT & Clinical & Administration \\ \midrule
    \multicolumn{1}{l|}{\multirow{2}{*}{Low (once per year)}}     & \multicolumn{1}{c|}{E} & 0.35 & 0.3      & 0.3            \\
    \multicolumn{1}{l|}{}                                         & \multicolumn{1}{c|}{C} & 1   & 30       & 10             \\ \midrule
    \multicolumn{1}{l|}{\multirow{2}{*}{Medium (twice per year)}} & \multicolumn{1}{c|}{E} & 0.6 & 0.5      & 0.5            \\
    \multicolumn{1}{l|}{}                                         & \multicolumn{1}{c|}{C} & 2   & 60       & 20             \\ \midrule
    \multicolumn{1}{l|}{\multirow{2}{*}{High (once per month)}}   & \multicolumn{1}{c|}{E} & 0.8 & 0.7      & 0.7            \\
    \multicolumn{1}{l|}{}                                         & \multicolumn{1}{c|}{C} & 12  & 360      & 120   \\ \bottomrule           
    \end{tabular}
    \vspace{0.3cm}
    \caption{Evaluation parameters for control CIS-17.4.}
    \end{center}
    \vspace{-0.9cm}
\end{table}

We have assumed a user group ratio of size 1:30:10 that loosely follows the corresponding breakdown of hospital workforce in the United States\footnote{\scriptsize{\url{https://www.bls.gov/oes/current/naics3\_622000.htm.}}}:
81,790 computer, information system and security managers and analysts; 2,437,540 healthcare practitioners; 737,750 receptionists, healthcare record information clerks and other office and administrative support staff. 

\begin{table}[h]
    \begin{center}
    \label{tab:cis176}
    \begin{tabular}{@{}lcccc@{}}
    \toprule
    Control level \textbackslash Role                             &                        & ICT & Clinical & Administration \\ \midrule
    \multicolumn{1}{l|}{\multirow{2}{*}{Low (Tests)}}     & \multicolumn{1}{c|}{E} & 0.25 & 0.2      & 0.2            \\
    \multicolumn{1}{l|}{}                                         & \multicolumn{1}{c|}{C} & 1   & 30       & 10             \\ \midrule
    \multicolumn{1}{l|}{\multirow{2}{*}{Medium (Videos)}} & \multicolumn{1}{c|}{E} & 0.7 & 0.6      & 0.6            \\
    \multicolumn{1}{l|}{}                                         & \multicolumn{1}{c|}{C} & 2   & 60       & 20             \\ \midrule
    \multicolumn{1}{l|}{\multirow{2}{*}{High (Games)}}   & \multicolumn{1}{c|}{E} & 0.6 & 0.5      & 0.5            \\ 
    \multicolumn{1}{l|}{}                                         & \multicolumn{1}{c|}{C} & 4  & 120      & 40  \\ \bottomrule
    \end{tabular}
    \vspace{0.3cm}
    \caption{Evaluation parameters for control CIS-17.6.}
    \end{center}
    \vspace{-1.5cm}
\end{table}

\subsubsection{Safeguards.}
As safeguards, we have considered a representative pair from the SANS institute's CIS-17 
group of critical 
security controls\footnote{\scriptsize{https://www.cisecurity.org/controls/implement-a-security-awareness-and-training-program/.}}: 
CIS-17.4 ``Update Awareness
Content Frequently'' and CIS-17.6 ``Train Workforce on Identifying Social Engineering 
Attacks''. All values used in this case study, for these two safeguards, are
presented in Tables 2 and 3.

For CIS-17.4, we set the frequency of completion of 
the updated training (once per year, twice per year, or once per month 
- i.e., 12 times per year) as the level of control. 
As indirect cost $C(j,i)$, we consider the total time spent in training 
by the employees 
in group $i$ at application level $j$ (in this case is \textit{frequency}), 
which is proportionate to the size of the group and the frequency of the training. 
This time can be translated to some financial cost (in \$) resulting from loss of productive 
working hours. In this way, the indirect cost can be subtracted from the expected loss 
comprising the final utility value of the Defender in each cell of the game utility matrix. 

For CIS-17.6, we set the nature of the work-based training 
(tests, videos, games) as the levels of control. Further, we set the corresponding efficacy values for each type roughly equivalent to their importance in helping predict user susceptibility 
to semantic social engineering attacks.
Specifically, \cite{heartfield2016you} has identified work-based security training with videos as 
the best predictor out of the three. In terms of efficacy values, we have differentiated slightly 
between groups based on our perceived rate of adoption of controls in each one. 
Specifically, we assume that adoption is greater for ICT than for clinical 
and administration employees. 
This is only for illustration purposes, so that the model can also take into account the group at each level of control.
We also assume, further, that the primary indirect cost is employee time required, with a ratio of 1:2:4 for the three control levels.

\subsection{Comparison with Alternative Defense Strategies}

In the following, we analyze the proposed model in two phases; 
(i) the game-theoretic; and (ii) the 0-1 Knapsack optimization. 
The \textit{first phase} evaluates different cybersecurity safeguard 
selection strategies using the utility table of the investigated Cybersecurity 
Safeguards Games (CSG) based on the use case discussed in the previous section. 
To evaluate our approach, we have created a simulated environment 
in Python which performs the attack sampling. For all comparisons performed, a sample 
size of 1,000 attacks was used. Such a sample is referred to as a \textit{run} in the results. 
In the following, we present the results, where 25 runs have been performed in each case and 
the average Defender Utility (in \$) seen across the runs have been plotted.

More specifically, we have simulated  
$\Gamma_{\sigma,2}$ and $\Gamma_{\sigma,3}$ (please see Table \ref{tab:list_symbols} for the notation) 
for the two different safeguards presented
in the use case, i.e., CIS 17.4 (denoted as $\sigma=1$) and 17.6 (denoted as $\sigma=2$).  
The games $\Gamma_{1,2}$, $\Gamma_{2,2}$ exhibit maximum safeguard application level of 2 (Medium), 
while the games $\Gamma_{1,3}$, $\Gamma_{2,3}$ are investigated up to application level 3 (High).
Each CSG generates a utility table that we use to derive three different Defender application
level selection strategies:
\begin{itemize}
    \item \textit{Nash} Safeguard Strategy (NSS), as described in Section 3 and computed using the
     the open source \emph{Nashpy} Python 
     library\footnote{\scriptsize{\url{https://nashpy.readthedocs.io/en/stable/index.html}}}. 
    \item the \emph{Weighted} Safeguard Strategy (WSS), which distributes the choice of a safeguard level 
    over the weighted expected utility of the CSG by computing 
    probability $\delta_{\sigma,j}$ of choosing application level $j$ of safeguard $\sigma$
    as follows: 
    \[
    \delta_{\sigma,j}:= \frac{\sum_{i=1}^{|\mathcal{U}|} U_d(j,i)}
    {\sum_{j=1}^{|\mathcal{L}|} \sum_{i=1}^{|\mathcal{U}|} U_d(j,i)}
    \]
    % \item \emph{Risky}: a profile always preferring the \textit{first level} of available safeguard
    \item the \emph{Cautious} Safeguard Strategy (CSS), which always prefers the 
    \textit{highest} application level of a safeguard.
\end{itemize}

\noindent Regarding adversarial strategies, we consider three profiles:
\begin{itemize}
    \item the \textit{Nash} Attacker who plays the Nash Attacking Strategy (NAS), presented in Section 3 and computed using the \emph{Nashpy} Python library.
    \item a \textit{Weighted} Attacker who plays the Weighted Attacking Strategy (WAS) by attacking a user group $i$ with probability $\frac{A_i}{\sum_{i\in \users} A_i}$, i.e. the Attacker attacks the different user groups proportionally based on the asset values they have access to.
    \item the \textit{Opportunistic} Attacker who uniformly chooses the different user groups to attack.
\end{itemize}

\begin{figure}[h]
     \centering
     \begin{subfigure}[b]{0.5\textwidth}
         \centering
         \includegraphics[width=\textwidth]{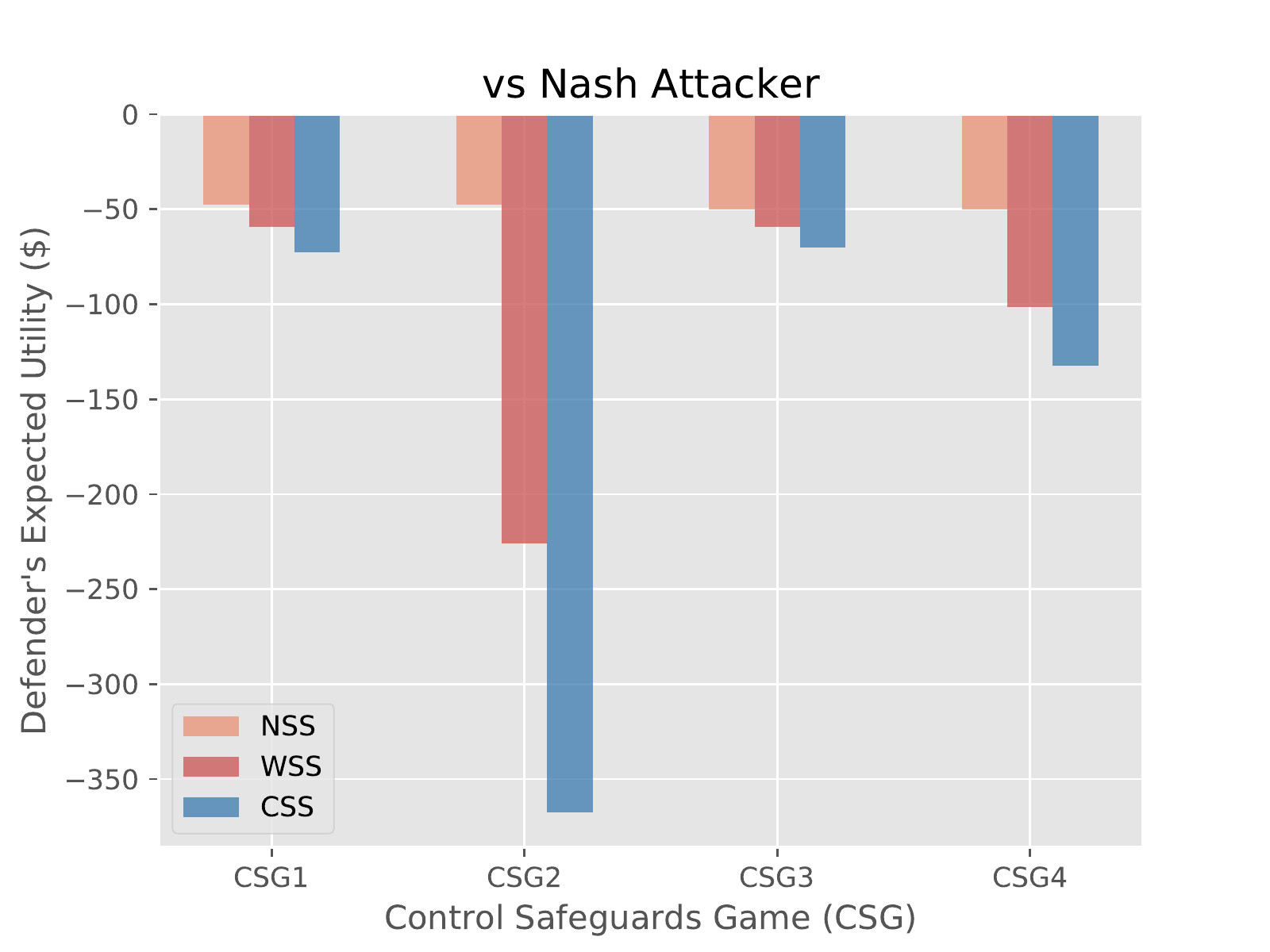}
         \caption{}
         \label{fig:utilites_against_nash_attacker}
     \end{subfigure}
     \hfill
     \begin{subfigure}[b]{0.49\textwidth}
         \centering
         \includegraphics[width=\textwidth]{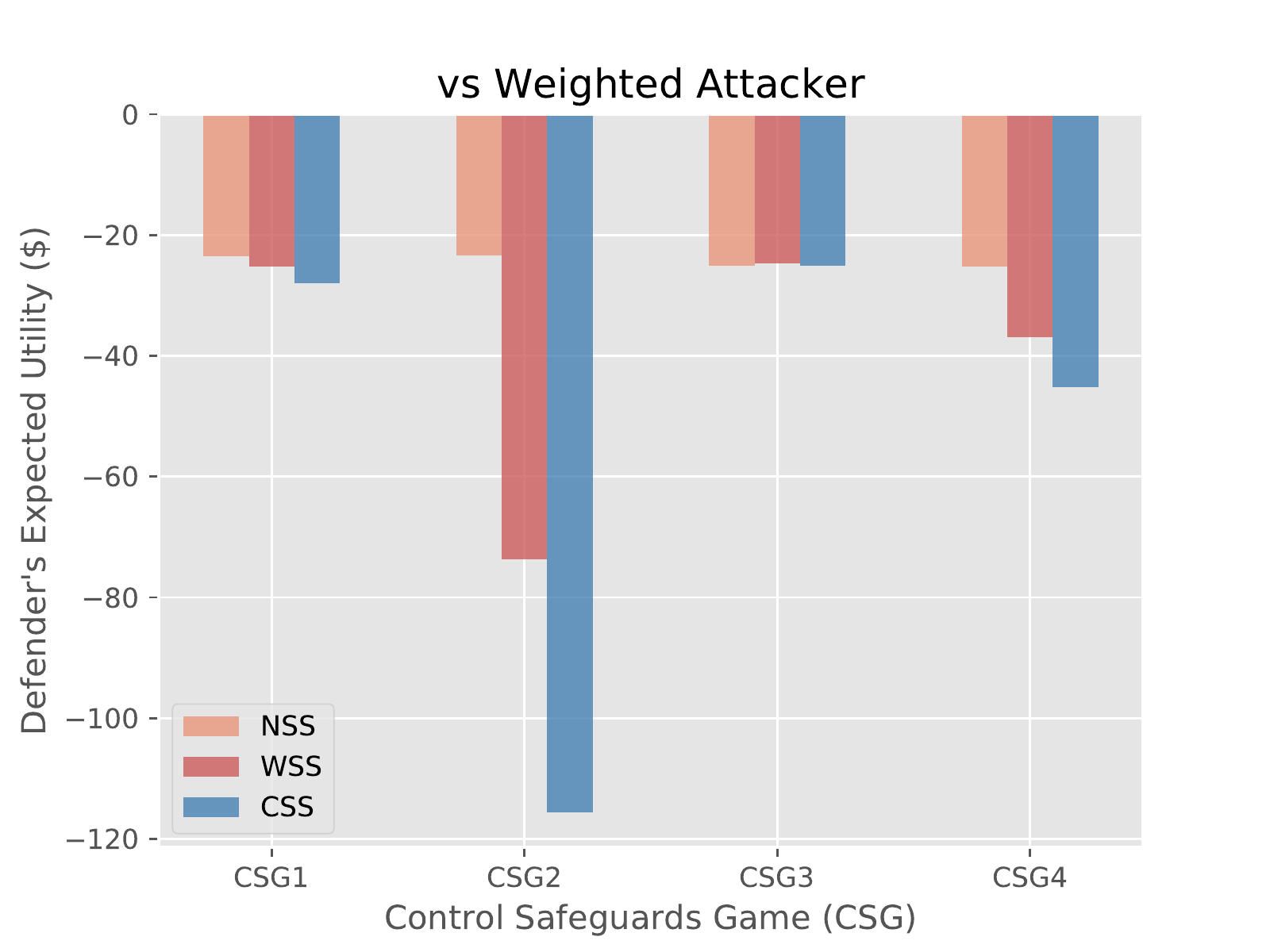}
         \caption{}
         \label{fig:utilities_against_weighted_attacker}
     \end{subfigure}
     \newline
     \begin{subfigure}[b]{0.5\textwidth}
         \centering
         \includegraphics[width=\textwidth]{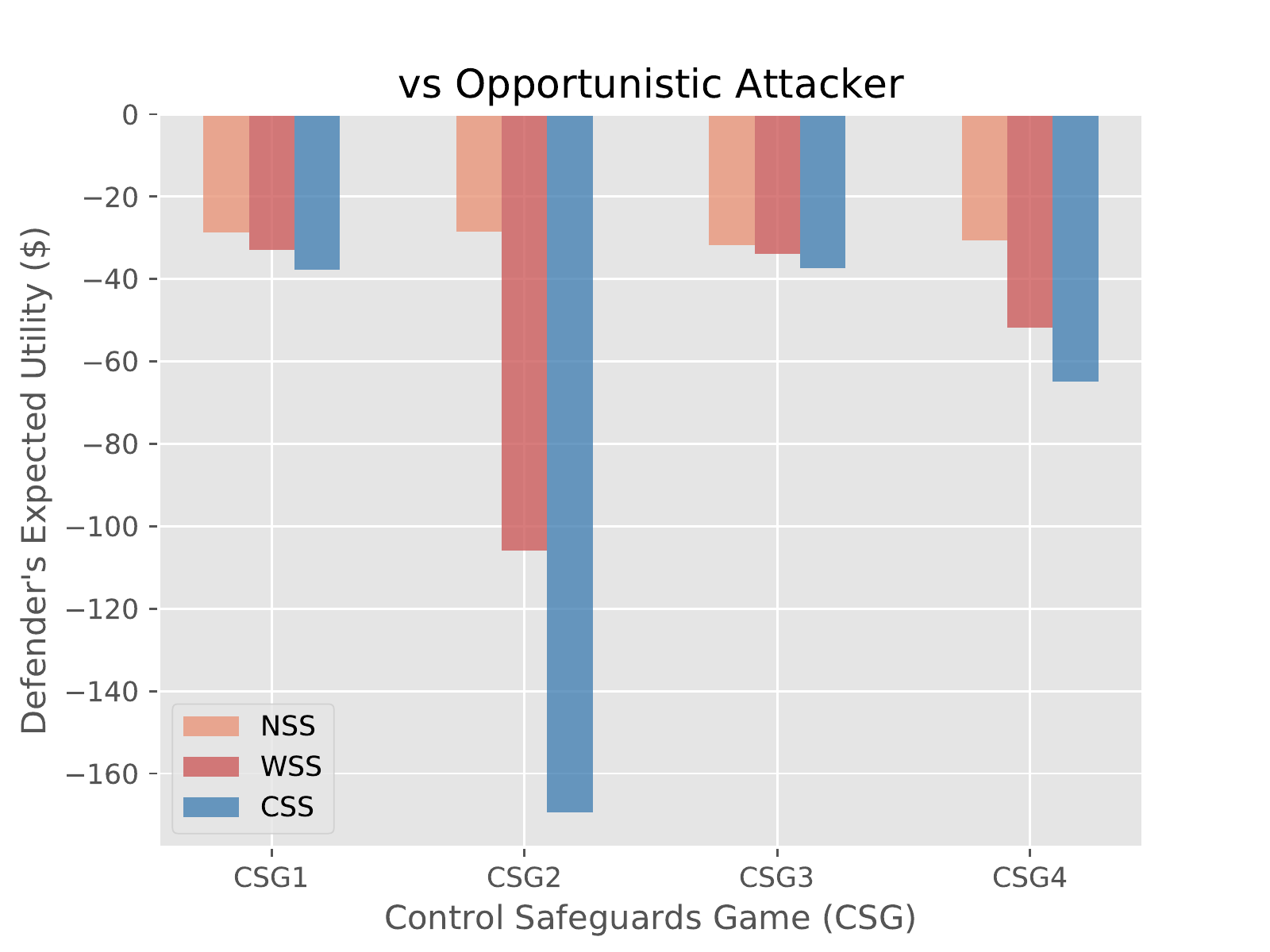}
         \caption{}
         \label{fig:utilities_against_opportunistic_attacker}
     \end{subfigure}
     \caption{Game-theoretic optimization results: Average Utility of the Defender over 1,000 attacks for 25 runs. 
     for various CSGs.}
    %  (a) Expected utility of defender profiles against Nash attacker for various 
    % CSGs. (b) Expected utility of defender profiles against Opportunistic attacker for various CSGs.
     \label{fig:defender_utility_against_attacker}
\end{figure}

Figure \ref{fig:defender_utility_against_attacker} illustrates the performance 
of NSS against WSS and CSS in terms of average Defender's utility over the 1,000 attacks
for 25 runs. In all cases, we contrast between Attackers who follow NAS and WAS. 

\subsubsection{Nash Attacker.}

The results, in Figure \ref{fig:defender_utility_against_attacker}(a), 
show that NSS outperforms both WSS and CSS when the Attacker chooses NAS. 
More specifically, the percentage improvement values, seen when choosing NSS, 
in comparison to WSS for the different games [$\Gamma_{1,2}, \Gamma_{1,3}, 
\Gamma_{2,2}, \Gamma_{2,3}$] are [20.2\%, 79.78\%, 16\%, 52.12\%], respectively. 
Likewise, when choosing NSS over CSS, we observe improvement values
of [34.48\%, 87.07\%, 28.57\%, 62.26\%] for the different games [$\Gamma_{1,2}, \Gamma_{1,3}, 
\Gamma_{2,2}, \Gamma_{2,3}$], respectively. 

\begin{remark}
    These results demonstrate an average improvement of approximately $42\%$ of 
    NSS over WSS and $53\%$ over CSS.  
\end{remark}

Comparably, the smallest average improvement for NSS over WSS is around 16\% when 
playing Control 17.6 at the maximum application level of 2 ($\lambda=2$). 
Likewise, the minimum improvement of NSS over CSS, approximately equal to 28\%, is 
for the same control and $\lambda=2$.
On the other hand, the maximum improvement seen in NSS over CSS is approximately 87\%, 
where the maximum improvement over CSS does not exceed 80\%, for Control 17.4.
and $\lambda=3$.
% \iMP{George could you imagine why? it has to do with the numbers chosen in the use case. any intuition
% could be good but not absolutely necessary.}

One of the primary reasons why naive-deterministic safeguard selection 
approaches perform poorly against the Nash Defending strategy is that they fail to 
incorporate the opponent's strategies. At the same time, we have considered CSG as a zero-sum game.
The class of zero-sum games offers a degree of freedom as it can be shown that assuming that 
the adversary's intentions are exactly opposite to the defender's assets, i.e., the Attacker 
seeks to cause maximum damage, any other incentive of the Attacker can only improve the 
Defender's situation \cite{rass2018password}.

\subsubsection{Weighted Attacker.}

When the Weighted Attacking Strategy is simulated, the results demonstrate that NSS has 
higher efficacy over WSS and CSS apart from one game $\Gamma_{2,2}$ in which both WSS and 
CSS perform approximately 2\% better 
than NSS (Figure \ref{fig:defender_utility_against_attacker}(b)). 
This difference is negligible making NSS
being at least as good as the rest of the Defending strategies in all investigated games. 
Despite the performance of NSS in $\Gamma_{2,2}$, for the rest of the games, NSS
performs significantly better than WSS and CSS. 
The percentage improvement values, seen when choosing NSS, 
in comparison to WSS and CSS for $[\Gamma_{1,2}, \Gamma_{1,3}, 
\Gamma_{2,2}, \Gamma_{2,3}]$ are [7.34\%, 70.25\%, -2.25\%, and 32.29\%] 
and [15.1\%, 80.44\%, -2.1\%, and 44.79\%], respectively. 
% \MP{George, can you think of any intuitive reason? or give any other
% connotation - no need to if it seems impossible}

\begin{remark}
    These results demonstrate an average improvement of approximately $28\%$ of 
    NSS over WSS and $34\%$ over CSS.  
\end{remark}

The smallest average improvements for NSS over WSS and CSS are approximately
7\% (in $\Gamma_{1,2}$) and 15\% ($\Gamma_{1,2}$), respectively, and the maximum 
average improvement values are 70\% (in $\Gamma_{1,3}$) and 80\% (in $\Gamma_{1,3}$).

\subsubsection{Opportunistic Attacker.}

Finally, when the Opportunistic Attacking Strategy is simulated, the results 
demonstrate that NSS has higher efficacy over WSS and CSS
(Figure \ref{fig:defender_utility_against_attacker}(c)). 
The percentage improvement values, seen when choosing NSS, 
in comparison to WSS and CSS for $[\Gamma_{1,2}, \Gamma_{1,3}, 
\Gamma_{2,2}, \Gamma_{2,3}]$ are [13.3\%, 74.24\%, 5.4\%, and 40.8\%] 
and [23.73\%, 83.4\%, 12.33\%, and 52.51\%], respectively. 

\begin{remark}
    These results demonstrate an average improvement of approximately $33\%$ of 
    NSS over WSS and $43\%$ over CSS.  
\end{remark}

The smallest average improvements for NSS over WSS and CSS are approximately
5\% (in $\Gamma_{2,2}$) and 12\% ($\Gamma_{2,2}$), respectively, and the maximum 
average improvement values are 74\% (in $\Gamma_{1,3}$) and 83\% (in $\Gamma_{1,3}$).

We notice that the highest improvements among the three different Attacking strategies
are introduced by the first scenario where Nash Attacker is simulated. 
This was anticipated as at the Nash Equilibrium the Defender does the best against 
a rational Attacker. Between the results for Weighted and Opportunistic Attacker, 
NSS is more efficient against an Opportunistic Attacker than a Weighted one.

\subsection{Analysis of the Investment Problem}

The Knapsack optimization phase investigates the optimal investment in Nash Safeguards Plans 
(NSPs) given a budget $B$ (for details refer to section \ref{sec:investment_NSP}). 
The Knapsack takes as input every NSP generated in the game-theoretic phase and recommends 
a single solution which minimizes the aggregated risk of all the user groups while satisfying
the investment budget constraint. This is different to the weakest-link model investigated by 
\cite{fielder2016decision}.
% \MP{George can we somehow justify this choice-there we had vulnerabilities, here wehave humans (i know they are vulnerabilities too) but can they mean something in terms of aggregated risks, 
% e.g. each of them is seen different from different Attackers, due to understanding of different human behaviorus,
% as opposed to technical vulnerabilities that do not have 
% behavioral characteristics?}
\begin{figure}[h]
     \centering
     \begin{subfigure}[b]{0.5\textwidth}
         \centering
         \includegraphics[width=\textwidth]{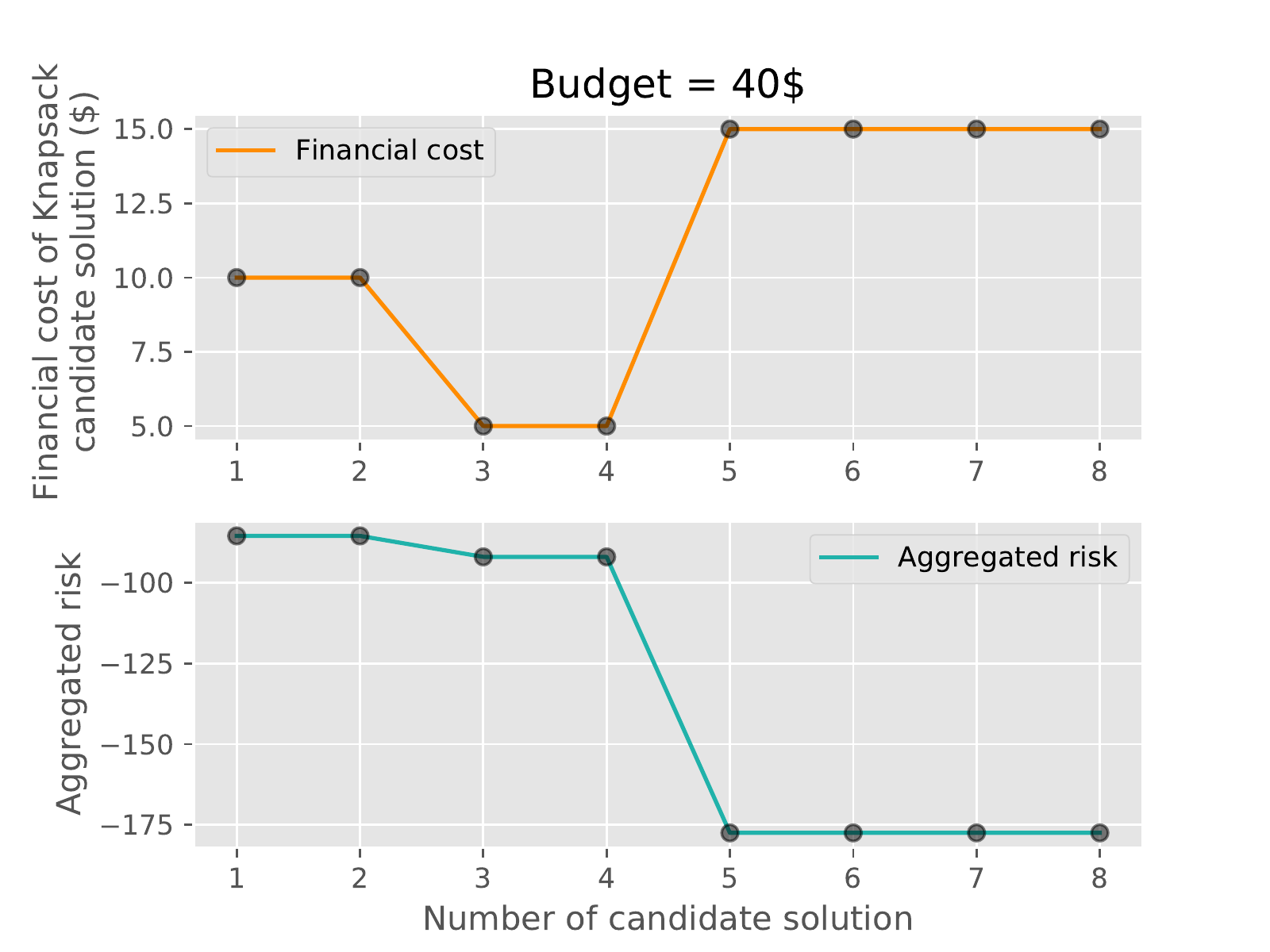}
         \caption{}
         \label{fig:cost_vs_risk_budget40}
     \end{subfigure}
     \hfill
     \begin{subfigure}[b]{0.49\textwidth}
         \centering
         \includegraphics[width=\textwidth]{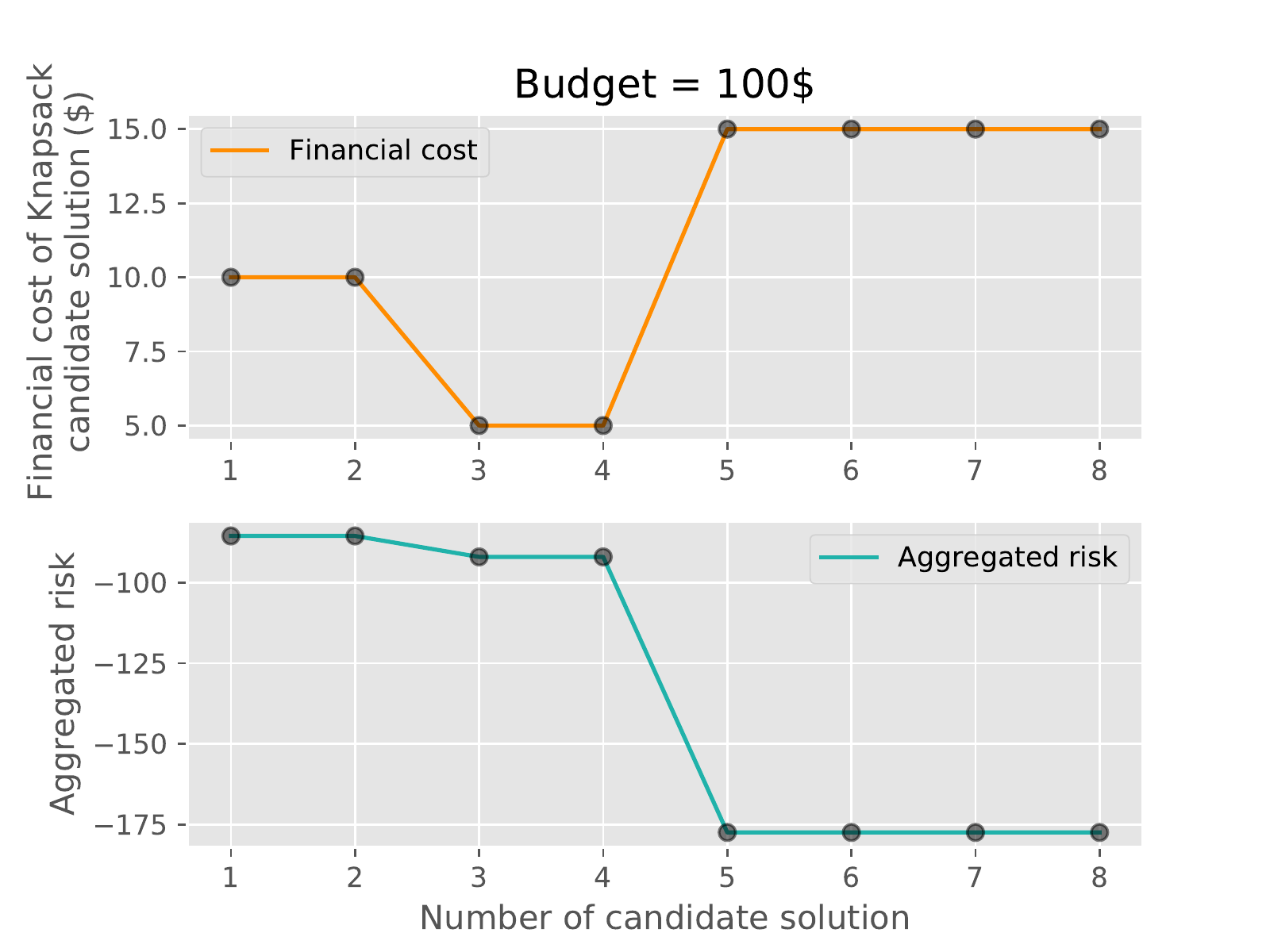}
         \caption{}
         \label{fig:cost_vs_risk_budget100}
     \end{subfigure}
     \caption{Knapsack selection over available candidate solutions.}
     \label{fig:cost_vs_risk}
\end{figure}

Figure \ref{fig:cost_vs_risk} presents the financial cost and aggregated risk 
overall users for each Knapsack candidate solution, i.e., a combination of NSPs
for two different available budget values.
We notice that there are multiple Knapsack optimal 
solutions, which are candidate solutions number $5,6,7$ and $8$. In the presence of 
multiple optimal solutions, the Knapsack solver, we have implemented, chooses the first option. 
For both budgets $40$ and $100$, the Knapsack optimization recommends investing in 
both CIS controls $17.4$ and $17.6$ at application level $1$ i.e., 
Low (once per year) and Low (Tests), respectively. Note here that the small size of 
the use case effectively prohibits high variability of the parametric values, 
which led to the selection of only two types.

% The formal results are presented as a snippet in the Supplementary section. 
% \iMP{Sakshyam shall we include the implemented algorithm? there are nice latex environment for this}

Note that the plots in Knapsack optimization only present the candidate solutions for the Nash Defender against Nash Attacker, in contrast to the plots in game-theoretic phase, 
(Figure \ref{fig:defender_utility_against_attacker}), which presents all three Defender strategies. This choice was made due to the Knapsack optimization not involving the notion of CSG. As a result of this, it does not optimize the overall indirect cost of safeguards when choosing NSPs which has been done in the previous phase. In addition, Knapsack does not consider the behavior of the Attacker characterizing all adversarial strategies as irrelevant to the Knapsack objective function.

% \newpage

\section{Conclusions}
In this paper, we have presented an approach, extending the previous work \cite{fielder2016decision}, which implements a cybersecurity safeguards selection model along with game-theoretic and Knapsack optimization tools. We have evaluated our model in a healthcare use case using the CIS group $17$ controls which attend to implementation of security awareness and training programs for employees. The simulation results demonstrate that the Nash Safeguard Strategy comfortably outperforms common-sense selection strategies, such as the Weighted and Cautious, in terms of Defender's expected utility over a large number of attacks. This work is our step towards integrating the developed Optimal Safeguards Tool (OST) within cybersecurity risk management and investment environments. 

An interesting extension to this work would be to capture the real-world uncertainty about an Attacker's type, for example considering a Bayesian game of application level selection. Furthermore, we plan to bring together several objective functions for Knapsack to compare the performance of the investment strategies. As the next steps, we aim at creating a use case with greater size of safeguards in collaboration with healthcare organizations. We also aim at using the well-known repository of cybersecurity safeguards like the 20 CIS controls or a list of Privacy Enhancing Technologies (PETs) to support our research.

\section{Acknowledgments} 
We thank the reviewers for their valuable feedback and comments.
\vspace{0.25cm}

\noindent Emmanouil Panaousis is partially supported by the European Commission as part of the CUREX project (H2020-SC1-FA-DTS-2018-1 under grant agreement No 826404). The work of Christos Laoudias has been partially supported by the CUREX project (under grant agreement No 826404), by the European Union's Horizon 2020 research and innovation programme (under grant agreement No 739551 (KIOS CoE)), and from the Republic of Cyprus through the Directorate General for European Programmes, Coordination and Development.

\begin{acronym}
% \acro{CVE}{Common Vulnerabilities and Exposures}%
% \acro{CVSS}{Common Vulnerability Scoring System}%
\end{acronym}

% Bibliography
\bibliographystyle{unsrt}
\bibliography{bibliography}

\begin{acronym}
  \acro{SME}{Small and Medium Enterprise}
  \acro{IoT}{Internet of Things}
\end{acronym}

% \appendix
% \newpage
% \section{Supplementary Content} \label{sec:supplementary}
% \begin{figure}[t]
%     \centering
%     \includegraphics[width=\textwidth]{plots/results_con1}
%     \caption{}
%     \label{fig:my_label}
% \end{figure}
% \begin{figure}[b]
%     \centering
%     \includegraphics[width=\textwidth]{plots/results_con2}
%     \caption{}
%     \label{fig:my_label}
% \end{figure}

\end{document}